\documentclass[12pt,aps]{revtex4}
\usepackage{graphicx}
\usepackage[sort&compress]{natbib}
\bibpunct{}{}{,}{s}{}{}

\usepackage[bf,small]{caption2} 
\usepackage{parskip}

\begin{document}

\clearpage
\title{The interface between silicon and a high-k oxide}
\author{Clemens J. F\"orst$^{*,\dagger}$,  Christopher R. Ashman$^{*}$, 
Karlheinz Schwarz$^{\dagger}$ \& Peter E. Bl\"ochl,$^{*}$}
\affiliation{$^*$ Clausthal University of Technology, Institute for
Theoretical Physics, Leibnizstr.10, D-38678 Clausthal-Zellerfeld,
Germany}
\affiliation{$^\dagger$ Vienna University of Technology, Institute for
Materials Chemistry, Getreidemarkt 9/165-TC, A-1060 Vienna, Austria}
\begin{abstract}
\end{abstract}
%
%\pacs{}
%
%\date{} 
%
\maketitle

\textbf{The ability to follow Moore's Law\cite{Moore} has been the
basis of the tremendous success of the semiconductor industry in the
past decades. To date, the greatest challenge for device scaling is
the required replacement of silicon dioxide-based gate oxides by
high-k oxides in transistors. Around 2010 high-k oxides are required
to have an atomically defined interface with silicon without any
interfacial SiO$_2$ layer. The first clean interface between silicon
and a high-K oxide has been demonstrated by McKee {\it \bf et al}.\cite{McKee98}
Nevertheless, the interfacial structure is still under debate.  Here
we report on first-principles calculations of the formation of the
interface between silicon and SrTiO$_3$ and its atomic structure.
Based on insights into how the chemical environment affects the
interface, a way to engineer seemingly intangible electrical
properties to meet technological requirements is outlined.  The
interface structure and its chemistry provide guidance for the
selection process of other high-k gate oxides and for controlling
their growth.  Our study also shows that atomic control of the
interfacial structure can dramatically improve the electronic
properties of the interface. The interface presented here serves as a
model for a variety of other interfaces between high-k oxides and
silicon. }

According to the International Technology Roadmap for
Semiconductors\cite{roadmap}, gate oxides shall be scaled to a
thickness below 1.5 nm as early as 2005. This corresponds to only a
few atomic layers.  Due to the quantum mechanical tunneling effect,
electrons can pass directly through such a thin layer of an insulating
material resulting in large static power consumption.  Since about
five years ago, intense research has been performed to replace silicon
dioxide related materials with other oxides that have a higher
dielectric constant (high-k).  This would allow the use of a thicker
gate oxide while retaining the electrical properties of an ultrathin
SiO$_2$ film.  Thus device miniaturisation could be continued without
suffering from tunneling currents through the gate oxide.

Interfacing a new oxide with silicon is a major challenge.  A gate
oxide has to fulfil a number of requirements: Not just intrinsic
properties such as the high dielectric constant and low defect
concentrations are important.  The gate oxide must also be chemically
stable in contact with silicon, it must have a sufficiently large
injection barrier, and it must not have interface states in the band
gap of silicon.  Oxides related to silicon dioxide, used in today's
transistors fulfil these requirements in a unique way due to their
strong bonds to silicon and their flexible bonding network. They fail
only due to their low dielectric constant.  Retaining the same
beneficial properties for high-k oxides has turned out to be very
difficult.  The first high-k oxides introduced technologically are
likely to be amorphous oxides or silicates of Hf and Zr with an interfacial
SiO$_2$ layer. Around 2010 however, an interfacial SiO$_2$ layer
will no longer be tolerable and oxides with an atomically
well defined interface with silicon will be required.

Following the first reports of epitaxial growth of AO and ABO$_3$
compounds\cite{Ishiwara88}$^{-}$\cite{alexe98} McKee {\it et
al}.\cite{McKee98}$^,$\cite{McKee93}$^{,}$\cite{McKee01} demonstrated in
1998 that an atomically abrupt crystalline interface between SrTiO$_3$
and silicon can be formed. However, the atomic structure proposed for
this interface is questionable because it exhibits silicon atoms with
coordinations rarely found elsewhere in nature.

Before discussing the formation of the interface we need to review the
clean (001) surface of silicon and describe its changes due to Sr
adsorption.  On the unreconstructed silicon surface the atoms form a
square array.  Due to a lack of upper bonding partners, each atom has
two singly occupied dangling bonds pointing out of the surface.  Pairs
of silicon atoms dimerise, using up one dangling bond per atom to form
the dimer bond. This is called the dimer row reconstruction.  A second
rearrangement leads to the so-called buckled dimer reconstruction: One
atom of each dimer lifts up and the other shifts down, resulting in a
``buckled'' dimer.  This buckling causes both electrons to localise in
the upper silicon atom of a dimer, whereas the other silicon atom with
the empty dangling bond prefers a more planar arrangement.  

By depositing various amounts of Sr atoms onto the Si(001) surface we
explored the structural complexity of Sr adlayers\cite{Ashman03}.
This enables us to attribute atomic structures to the periodicities
observed experimentally\cite{Fan90}. Here we only summarise the
findings relevant for the present topic: Initially each Sr atom
donates two electrons into the empty dangling bonds of the surface. As
Sr is added, the dimer buckling vanishes since both dangling bonds of
a Si-dimer become filled with electrons.  Similar to
Ba\cite{Yao99,Wang99}, Sr first occupies the trough between the dimer
rows, in the centre of four dimers.  At a coverage of 1/2 monolayer
(ML) all positions in the trough are occupied and each dimer dangling
bond is filled with two electrons.  This $(2\times1)$ structure is the
only Sr-covered surface without surface states in the band gap of
silicon. Therefore, it is a suitable building block for an interface
without states in the gap, as required for device applications. The
finding of a sizeable band gap also explains why this surface is
fairly resistant to oxidation\cite{Liang01}.

The important finding from what has been discussed so far is that the
substrate surface can be chemically saturated by half a ML of Sr. Such
a surface is ``isoelectronic'' to an H-terminated silicon
surface. Hydrogen is known to be very effective at passivating
silicon. In the following, we therefore refer to the Si surface
covered with 1/2 ML of Sr as the Sr-passivated substrate.

\begin{figure}[h]
\centering
\includegraphics[width=0.70\textwidth,draft=false]{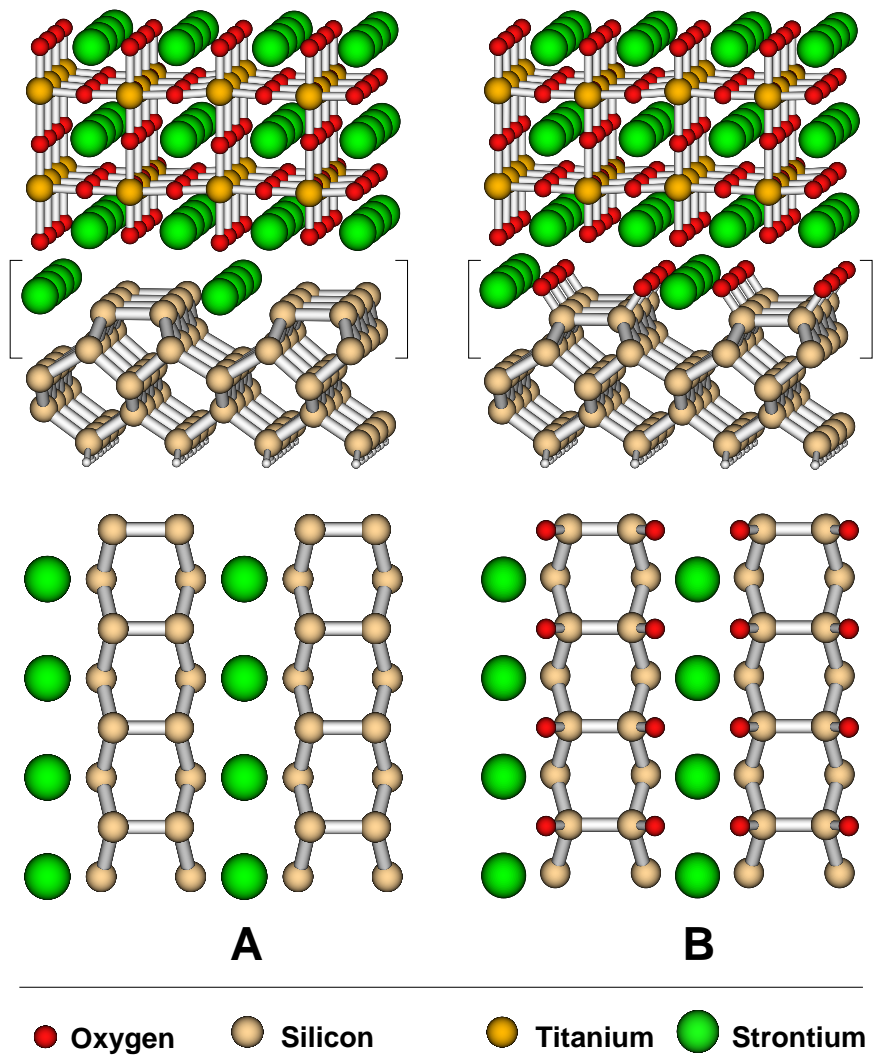}
\caption{Atomic structures of the SrTiO$_3$/Si(001) interfaces:
Left: unoxidised interface denoted in the text by \textbf{A}; Right:
oxidised interface denoted by \textbf{B}. Top: view slightly off the
[110] direction of silicon, which is parallel to the [100] direction
of the oxide. The topmost layer corresponds to the oxide surface of
our slab calculation. Bottom: view along the interface normal of the
interface layer.}
\label{fig1}
\end{figure}

After having gained insight into the metal overlayers, we investigated
the formation of an oxide layer.  We start from the Sr-passivated
substrate and simulate the deposition of one layer of SrO. During a
heating cycle to 600 K this single oxide layer reconstructs
significantly.  However, after placing two or more layers of SrO or
SrTiO$_3$ on top of the reconstructed SrO layer, the oxide layers
crystallise into their perfect bulk structure. Thus we obtain an
atomically abrupt interface between the silicon substrate and the
high-k oxide.  This interface structure, denoted by \textbf{A} and
shown in Fig.~\ref{fig1}, corresponds to the Sr-passivated silicon
surface matched to the nonpolar SrO layer of the oxide.

In addition to the chemical saturation of the substrate surface with
1/2 ML of Sr, the second important ingredient is the matching of the
charge patterns of the oxide and the Sr-passivated substrate surfaces
joined at the interface.

Whereas the SiO$_2$/Si interface relies on strong covalent bonds
across the interface and a flexible bond network of the oxide, the
interface described here is based on the chemical saturation of the
silicon surface with an alkaline earth metal, so that a template for
the deposition of a matching oxide is obtained.  

In a device the interface is exposed to a number of chemical
influences that affect the stability of the stack.  The most critical
question is the stability of the interface with respect to
oxidation. Oxygen ions can diffuse out of the gate oxide to the
interface.

In order to explore how the interface changes upon oxidation, we have
added oxygen atoms to a wide range of different sites. Oxygen first
attacks the surface silicon atoms at their vacant coordination
sites. After introducing 1 ML of oxygen into the interface all these
sites are consumed. The resulting structure, in the following denoted
by \textbf{B}, is shown in Fig.~\ref{fig1}.  Additional oxygen atoms,
up to a total oxygen content of 1.5 ML, insert into the dimer
bonds. As explained below, structure \textbf{B} and the dimer-oxidised
variant of structure \textbf{B} are the optimum choices for device
applications.

\begin{figure}[h]
\centering
\includegraphics[width=0.5\textwidth]{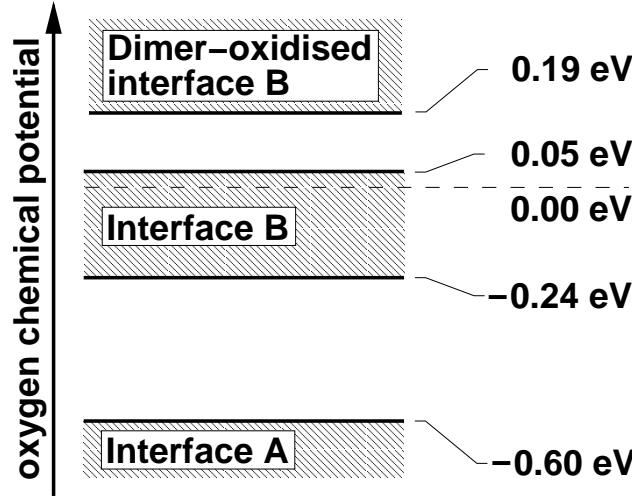}
\caption{ Phase diagram for interface oxidation: Shaded areas
indicate the stability regions of the defect-free interfaces
\textbf{A}, \textbf{B} and the dimer-oxidised variant of interface
\textbf{B}.  The blank regions separating them correspond to
disordered structures with an oxygen content that increases with
increasing chemical potential.  The external parameter is the oxygen
chemical potential.  The zero of the chemical potential corresponds to
the coexistence of bulk Si and SiO$_2$ ($\alpha$-quartz) in thermal
equilibrium.  }
\label{fig:oxidation}
\end{figure}

Fig.~\ref{fig:oxidation} illustrates the phase boundaries of the
interfaces as a function of the oxygen chemical potential. The chemical
potential is defined as the energy required to add a single oxygen
atom to the system.  It is the driving force for oxidation, which can
be controlled externally, for example, by choosing the appropriate
temperature and oxygen partial pressure during growth.  In thermal
equilibrium the chemical potential is related to the partial pressure
and temperature of the growth chamber. It should be noted, however,
that the formation of the stack mostly involves non-equilibrium
processes, so that the chemical potential at the interface lags behind
the value reached in the growth chamber. The oxygen chemical potential
needs to be sufficiently low to avoid formation of an interfacial
SiO$_2$ layer but high enough to avoid a large oxygen vacancy
concentration in the gate oxide, that may cause trap assisted leakage
currents.

Our results indicate that structures \textbf{A} and \textbf{B} can be
formed selectively in the absence of an interfacial SiO$_2$ layer. The
processing window defined as the stability region of interface
\textbf{B} extends from $-0.24$~eV to $-0.05$~eV and corresponds to a
range of partial pressures of nearly three orders of magnitude at
1000~K. This region lies entirely below the coexistence line of Si and
SiO$_2$.

In addition, we find indications of sizeable thin film effects, which
delay the formation of an interfacial SiO$_2$ layer. This can be
inferred from the fact that the oxidation of all dimer bonds requires
a chemical potential of 0.19~eV, and subsurface oxidation starting
from interfaces \textbf{A} or \textbf{B} require energies in excess of
0.91~eV above the coexistence line of Si and SiO$_2$.

The interface structure proposed in this work is quite different and
much simpler than previously assumed: It was believed that an
interfacial silicide\cite{McKee98,McKee93} or
silicate\cite{Wang01,Droopad01} layer must be formed. Such a layer is
not present in the interface that emerged from our simulations.
Nevertheless, our simulated SrTiO$_3$/Si(001) interfaces reproduce the
undisputed features of the Z-contrast images of McKee {\it et
al}.\cite{McKee98} such as the pattern of interfacial Sr atoms and the
oxide-substrate registry.  Thus the lateral alignment of the columns of
Sr and Ti atoms relative to the Si substrate and the $(2\times 1)$
periodicity can clearly be identified.

We also performed calculations on the interfaces proposed by McKee {\it
et al}.\cite{McKee98, McKee93} and by Wang {\it et al}.\cite{Wang01,Droopad01}. As also
shown by Robertson {\it et al}.\cite{robertson02}, the interface proposed by
McKee {\it et al}.\cite{McKee98, McKee93} is metallic, which is detrimental for
device applications. The same applies for the structural proposal of
Wang {\it et al}.\cite{Wang01,Droopad01}, which reconstructs significantly upon
relaxation.

Based on electron-count arguments, Robertson {\it et al}.  recently proposed a
structure\cite{robertson02} that is related to our dimer-oxidised
variant of structure \textbf{B}. It differs in that the oxide starts
with the TiO$_2$ layer instead of a SrO layer and it is derived from a
$c(2\times2)$ dimer reconstruction of the silicon surface.  If we modify
our interface by terminating the oxide with a TiO$_2$ layer instead of a
SrO layer, it is more stable than Robertson's proposal by 0.19~eV per
$(1\times1)$ surface unit cell.  Having an oxide terminated by SrO is,
however, favourable compared to TiO$_2$ terminated oxides, because the
TiO$_2$ layer and the substrate in direct contact are expected to react,
as pointed out by Robertson {\it et al}.\cite{robertson02}.  Conceptually
similar interfaces with a TiO$_2$ interface layer have been investigated
by Zhang {\it et al}.\cite{zhang03} who rule out their use in devices on the
basis of their electronic properties.

A critical parameter for gate stacks is the injection barrier, which
is the offset between the conduction band edges of the silicon
substrate and the oxide.  It prevents electrons from entering the
oxide conduction band, where they can cross the gate oxide. For device
applications the injection barrier should be larger than 1
eV\cite{roadmap}. There are indications\cite{Chambers01,Robertson00}
that the injection barriers for most high-k oxides are too low.

Before we discuss our results on the band offsets we need to briefly
touch upon the band gap problem of density functional theory
(DFT)\cite{Kohn,KohnSham}: Typically the one-particle energies
obtained in these calculations underestimate the band gap. Therefore,
there is an uncertainty in our calculated injection barriers.
Assuming that the error of the valence band edge is negligible as
required in exact DFT, and using the experimental band gaps of silicon
and SrTiO$_3$\cite{Noland54}, we anticipate that our calculations
underestimate the injection barrier by 0.7--0.8~eV.  

For the interface \textbf{A} we obtain an injection barrier that is negative by
0.6~eV. Including our correction we estimate the injection barrier to
lie at 0.1--0.2~eV. The injection barrier lies below the
technologically required minimum. 

For the interface \textbf{B}, however, we obtain a positive injection barrier
of 0.5~eV. Adding the correction, our final estimate yields a positive
injection barrier of 1.2--1.3 eV, which fulfils the criterion.  The
margin is sufficiently large that a reasonable error in the
band-offset correction does not lead to an unacceptably low injection
barrier. Most important is the possibility to influence the injection
barrier by carefully choosing the processing conditions.

Note that band-offsets are frequently derived from properties of the
two bulk materials alone\cite{Robertson00}, disregarding the interface
structure and composition in detail. The interface between silicon and
SrTiO$_3$ is an example where the band offset can be engineered by
controlling the chemical environment.  The change of the band offset
due to oxidation is about 1.1~eV, and thus sizeable.  It results from
a dipole created when the electrons are transferred from the filled
dangling bonds of the surface silicon atoms to the oxygen atoms that
attach to the vacant coordination sites.

The injection barrier of interface \textbf{B} is fairly insensitive to
additional oxidation of the dimer bonds.  
An amorphous interfacial SiO$_2$ layer is likely to destroy these
restrictions and thus lead to a lower injection barrier.  

\section*{Methods}

We performed state-of-the-art electronic structure calculations and
ab-initio molecular dynamics simulations\cite{Car85} based on density
functional theory \cite{Kohn, KohnSham,PBE} and the projector
augmented wave method\cite{PAW94}. 

The calculations have been done on 5-layer slabs of silicon. The slab
calculations included a vacuum region of at least 6~\AA\ between
repeated slabs. The relevant calculations of the interfaces are done
in a $(2\times2)$ supercell.  All structures are relaxed
without symmetry constraints.  The hydrogen-terminated silicon back
plane has been kept frozen.

Our calculations used a plane wave cutoff of 30~Ry for the plane wave
part of the wave function. The frozen core approximation was
employed. Semi-core states of Sr and Ti have, however, been treated as
valence electrons. We used the following sets of projector functions
per angular momentum: 2s2p1d for oxygen, 2s2p1d for silicon, 3s2p2d
for strontium and 2s2p2d for titanium. 

For all calculations of Sr adsorption we used a grid with about 64
lateral k-points per ($1\times1$) surface unit cell. For the interfaces we
used a grid corresponding to 16 k-points per $(1\times1)$ unit
cell. For metallic structures we used the Mermin functional with a
temperature of 1000\,K and the extrapolation to zero Kelvin proposed
by Gillan\cite{gillan}.

The band offsets have been derived by relating the plane wave part of
the potential, averaged laterally, to band edges. The relative
displacement between potential and band edges has been obtained from
the epitaxially strained bulk materials.

The phase diagram in Fig.~\ref{fig:oxidation} was obtained from the
total energies as a function of the oxygen chemical potential.  A number
of different, stoichiometric and non-stoichiometric, interface
structures with varying oxygen content have been considered.  The
regions where one of the stoichiometric interfaces (\textbf{A},
\textbf{B} or the dimer-oxidised variant of \textbf{B}) is most stable
are shaded.  Tests for the upper boundary of interface A with larger,
i.e. $(4\times4)$, unit cells confirmed that the phase boundaries
obtained in $(2\times2)$ unit cell are reliable to about 0.02~eV.

%====================================================================
\section*{Acknowledgements}
%====================================================================

We thank S.~Chambers, M.~Chisholm,  W.~Daum, A.~Dimoulas, J.~Fompeyrine, J.-P.~Loquet,
 R.A.~McKee, G.~Norga, and S.~Stemmer for useful discussions.  This
 work has been funded by the European Commission in the project
 "INVEST" (Integration of Very High-K Dielectrics with CMOS
 Technology) and by the AURORA project of the Austrian Science Fund.
 Parts of the calculations have been performed on the Computers of the
 ``Norddeutscher Verbund f\"ur Hoch- und H\"ochstleistungsrechnen
 (HLRN)''.

\vspace*{1cm}

Correspondence and requests for materials should be addressed to 
P.E.B.(email: Peter.Bloechl@tu-clausthal.de).

%
%====================================================================

%====================================================================

%
\end{document}